# Weak polarized lines of the second solar spectrum (Na, Al, H, He, Ti, Li, Sc, C$_2$ and MgH) observed at the Pic du Midi turret dome

*Jean-Marie Malherbe (emeritus astronomer)*

Observatoire de Paris, PSL Research University, LIRA, France

Email: Jean-Marie.Malherbe@obspm.fr; ORCID: https://orcid.org/0000-0002-4180-3729

25 March 2026

**ABSTRACT**

The second solar spectrum is the solar spectrum of Stokes parameter Q observed in linear polarization close to the solar limb. It differs significantly of the usual intensity spectrum (Stokes parameter I). The second solar spectrum contains in the visible range a few polarized lines with Q/I over 1% (such as CaI, SrI, SrII, BaII), but most lines exhibit weak or very faint polarization rates (Q/I < 0.3%). This paper presents unpublished observations made in 2004-2006 of weak polarized lines performed with the Pic du Midi Turret Dome spectropolarimeter, such as atomic lines of Na, Al, H, He, Ti, Li, Sc as well as C2 and MgH molecules.

**I - Introduction**

Scattering processes on the Sun are sources of polarization in the solar spectrum (Stenflo & Keller, 1997). For reasons of symmetry, the polarization is zero at disk centre (symmetry of the radiation field around the line of sight) and increases towards the solar limb (where the line of sight is orthogonal to the incident radiation field of the photosphere). The Stokes parameter Q represents linear polarization parallel to the limb, and the ratio Q/I is the polarization rate. To determine this polarization, the slit of the solar spectrograph is placed close to and parallel to the limb in the vicinity of the solar north or south, in order to avoid magnetized regions, and the camera integrates for a long time to reach sufficient signal to noise ratio (SNR). Indeed, the polarization rate Q/I due to the anisotropic scattering is small: a few lines exhibit about 1%, such as SrI, SrII, BaII or CaI lines in the blue part of the spectrum. On the contrary, many lines present weak polarizations below 0.3%, which requires a high polarimetric sensitivity to resolve the Q/I profiles as a function of wavelength. Hence anomalous or unexplained faint polarizations were discovered by Stenflo et al (2000a). Figure 1 shows the second solar spectrum at 5" below the limb obtained by Gandorfer (2000, 2002, 2005) at Locarno with the Zürich IMaging POLarimeter (ZIMPOL) from 4500 to 6800 Angström ; this range contains most of the lines we observed at Pic du Midi. The figure saturates for Q/I = 0.5%. Figure 2 shows a plot of Q/I as a function of wavelength with 0 < Q/I < 0.3%. Both figures reveal that the polarization spectrum considerably differs from the intensity spectrum and that above 4500 A, the polarization rate Q/I is weak. As a consequence, instrumental polarization can be an obstacle for such observations, but in our case, we used the 50 cm refractor of the Pic du Midi turret dome (equatorial mount, Roudier *et al*, 2021) with the polarimeter (Malherbe *et al*, 2007) at the primary focus, without any reflection, which guarantees almost no parasitic polarization. After passing through the polarimeter, the beam is injected into the 8 m spectrograph (Mouradian, 1980), a Littrow design of 8 m focal length. The dispersive element is a grating which offers 10 mA spectral resolution in the blue part of the spectrum. The CCD is an interline chip with electronic shutter running at about 5 -10 Hz.

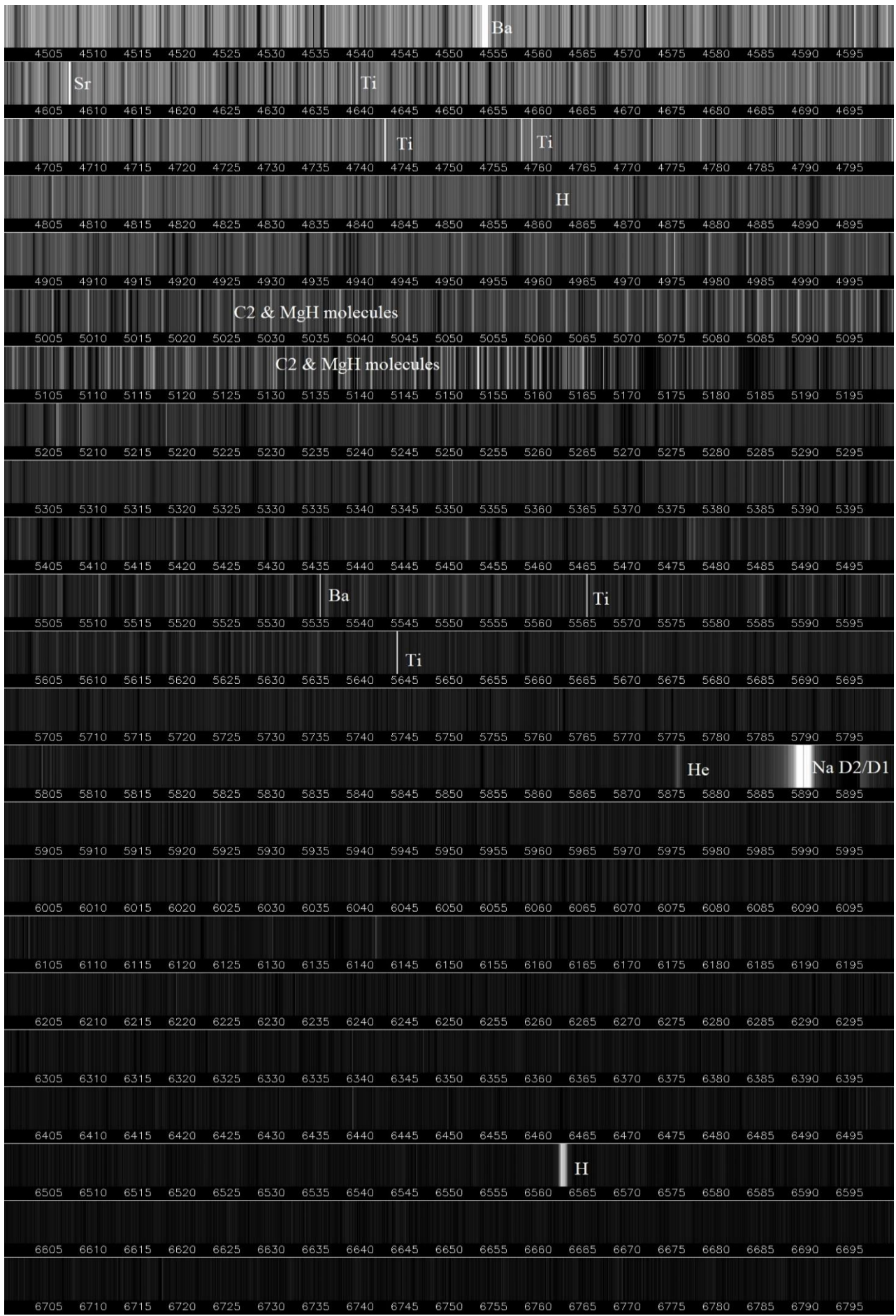

*Figure 1: the second solar spectrum (Q/I) in the range 4500-6800 A (after Gandorfer, 2000,2002,2005)*

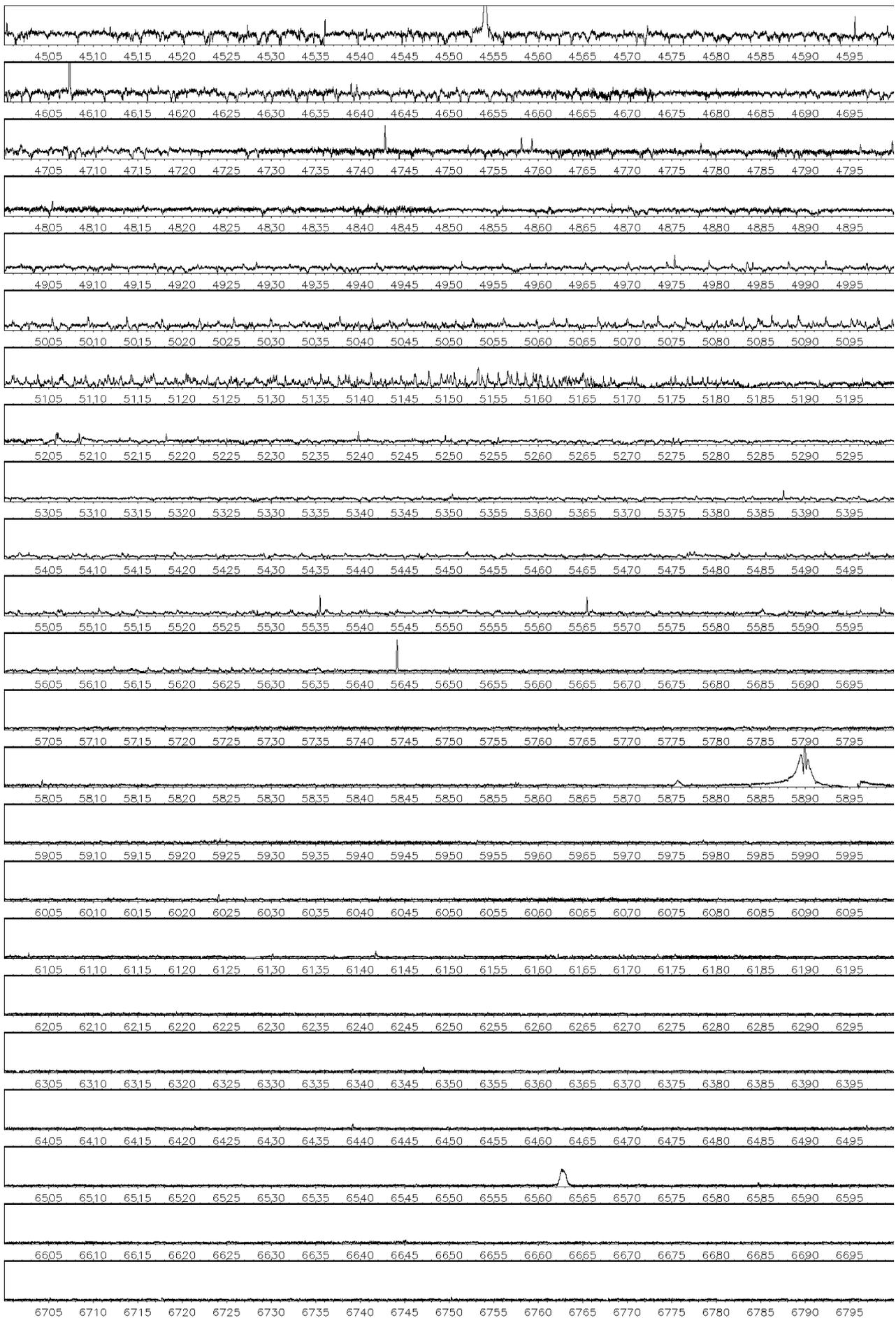

*Figure 2: the second solar spectrum (Q/I) in the range 4500-6800 A (after Gandorfer, 2000,2002,2005)*

## II - Observations with the Pic du Midi spectropolarimeter

The polarimeter is shown by Figure 3. It is located at the F1 focus of the refractor at F/13 and protected by an UV/IR cutoff filter (bandpass 380-700 nm). It is made of two Liquid Crystal Variable Retarders (LCVR) and a precision dichroic linear polarizer from Meadowlark company. Then the beam is injected into the spectrograph by a flat mirror at 45°, before magnification by a lens (equivalent focal length from 30 m to 60 m according to the magnification). The polarimeter can run at 50 Hz but is limited by the speed of the camera and exposure time. It has no chromatism, because the voltage is adapted to the observed line to produce exactly quarter or half wave retardance. Figure 4 shows the theoretical calibration curve of the two waveplates (theoretical retardance as a function of voltage) and our experimental calibration with narrow band filters.

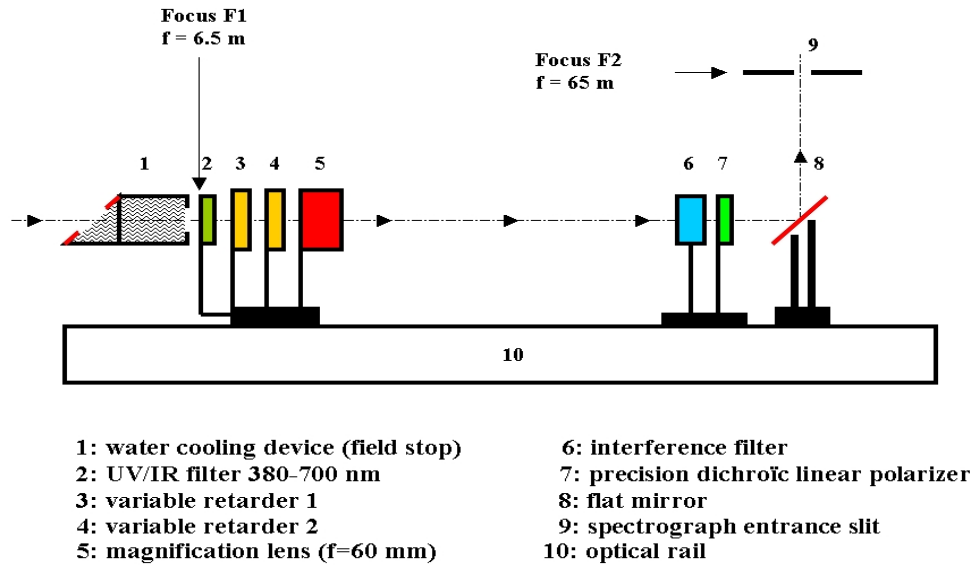

1: water cooling device (field stop)
2: UV/IR filter 380-700 nm
3: variable retarder 1
4: variable retarder 2
5: magnification lens (f=60 mm)
6: interference filter
7: precision dichroïc linear polarizer
8: flat mirror
9: spectrograph entrance slit
10: optical rail

*Figure 3: the liquid crystal polarimeter at the primary focus of the refractor*

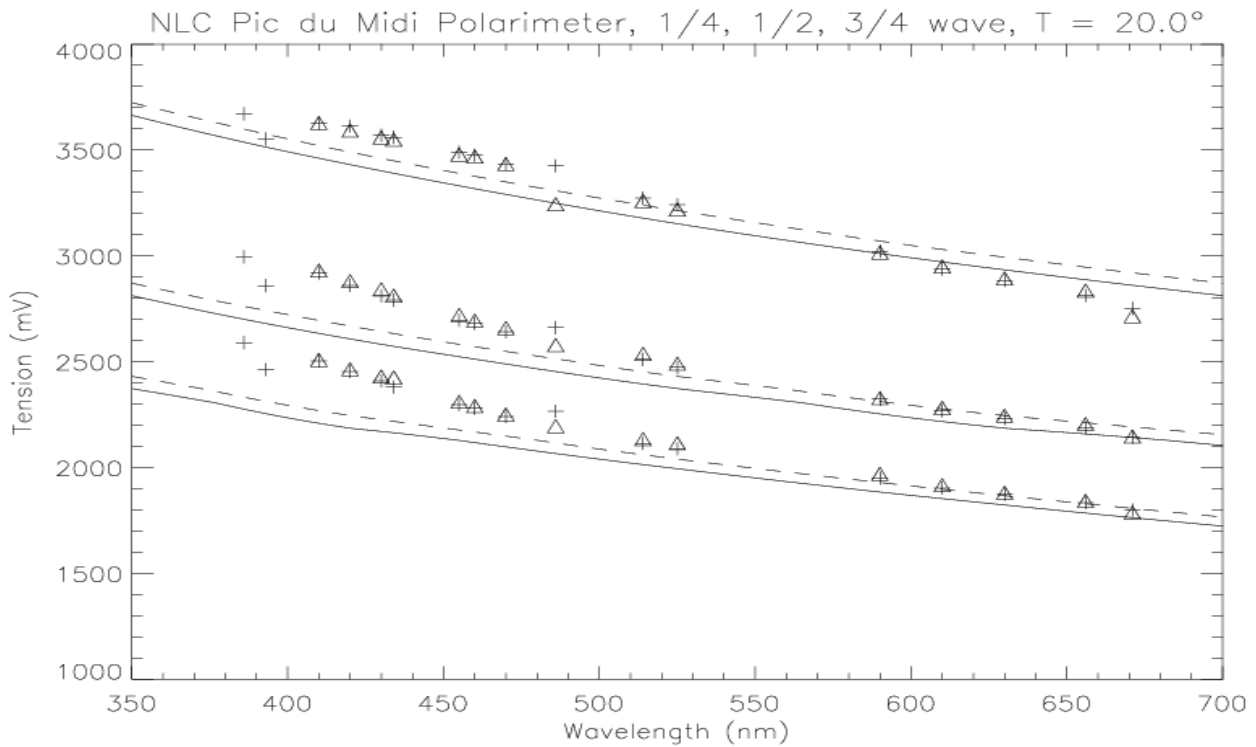

*Figure 4: the theoretical calibration curves (voltage) for quarter, half and three quarter wave, as a function of wavelength (nm), and our experimental measurements (symbols) for the two LCVR*

The polarimeter delivers the following combination S of Stokes parameters (I, Q, U, V):

$$S = (1/2) ( I + Q \cos\delta_2 + \sin\delta_2 (U \sin\delta_1 - V \cos\delta_1) )$$

where $\delta_1$ and $\delta_2$ are the retardances. In our observation, $\delta_1 = 0$ and $\delta_2$ modulates between 0 and $\pi$ in order to give alternatively, at 5-10 Hz cadence, the combinations I+Q and I-Q in sequence. The camera is a classical interline CCD camera with micro lenses (SONY sensor with 6.5 microns pixel) and electronical shutter. As the full well capacity is only 18000 electrons, corresponding to 12 bits (4.5 electrons/count), we alternate at 10 Hz cadence 16 exposures of I+Q and I-Q which are added in real time before file writing. We register consequently 16 bit data corresponding to a full well capacity of 288 000 electrons, which corresponds to a signal to noise ratio (SNR) of about 500. Several hundreds of observations obtained within a loop were then summed in order to reach a final SNR of about 10000 for typically 400 accumulations of 16 bit spectra. The SNR was then increased to reach 100000 by summing pixels along the slit (maximum 800 pixels). This allows to evidence polarization rates Q/I as low as $10^{-5}$. However, our observations present some limitations (for instance I+Q and I-Q measurements are sequential) and some optical defaults (some of them were partly corrected, such as field non-uniformity or interference fringes; fringes come from the interference filters which serve for order selection, they are more intense in the red than in the blue). In practice, the final polarization sensitivity of our experiment lies in the range $10^{-5}$ to $10^{-4}$ depending on the number of accumulated frames and pixel summation. The spectral resolution is around 10 mA, depending on the line (profiles are more resolved in the blue). The spectral pixel along the slit is 0.2 arcsec, and the length of the slit is 160". In our dataset, the continuum polarization is set to zero as our experiment does not allow to measure it. The dataset of FITS files is available on line at https://doi.org/10.57745/FEXIZ0, it contains high polarimetric sensitivity data obtained at about 10" of the solar limb ($\mu = \cos\theta = 0.15$) with the slit parallel to the limb. However, some data were got with the slit orthogonal to the limb. The polarimetric sensitivity of the dataset is greater than the one of Gandorfer's atlas, because we focused on narrow spectral domains with long exposure times.

**II – 1 - Aluminium Al 6697 A lines (figure 5)**

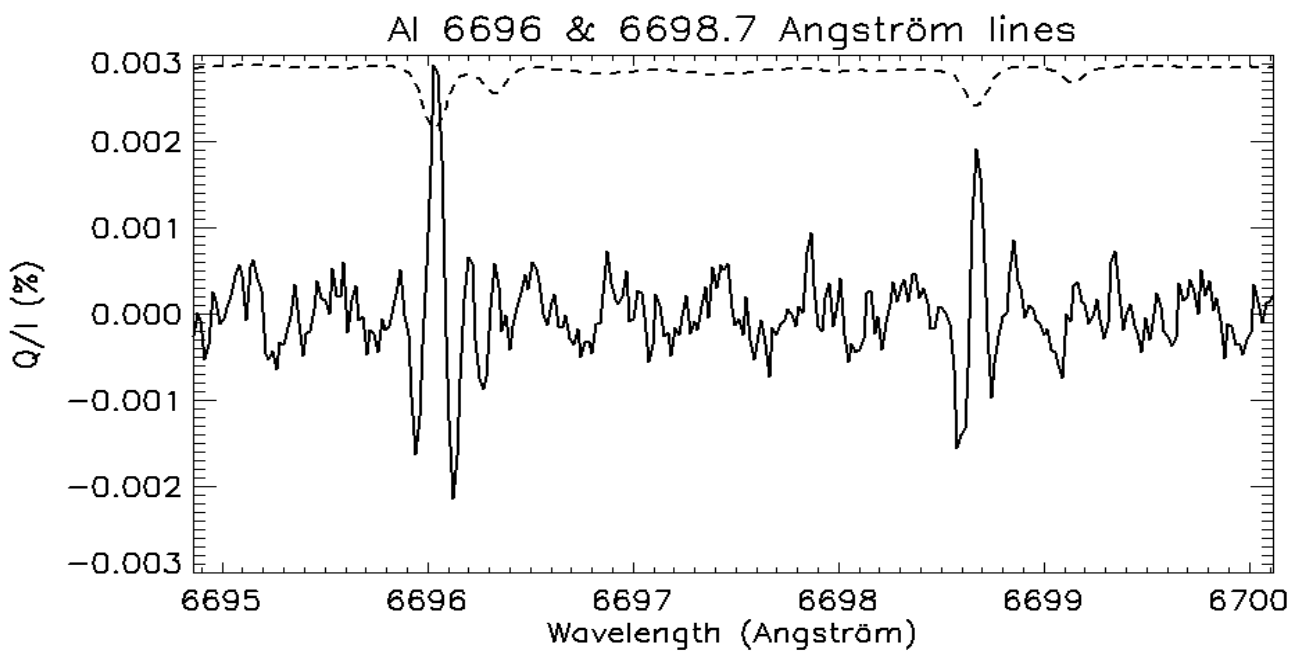

*Figure 5: Al lines at 6696 and 6698.7 A*

The polarization of Al lines (figure 5) was difficult to evidence by our experiment, as it is faint and perturbed by interference fringes of the selecting order filter, which were only partly removed.

**II – 2 - $C_2$ & MgH molecules: many lines around 5140 A (figures 6 & 7)**

The Hanle effect (depolarization and rotation of the plane of polarization) is operating in the presence of weak, turbulent and unresolved magnetic fieldq. It was studied by Berdyugina & Fluri (2004) for molecular lines. The large number of $C_2$ and MgH lines allows to use the differential effect between many lines of the same molecule to determine magnetic fields. Figures 6 & 7 show our observations of $C_2$ and MgH polarization near solar north and south.

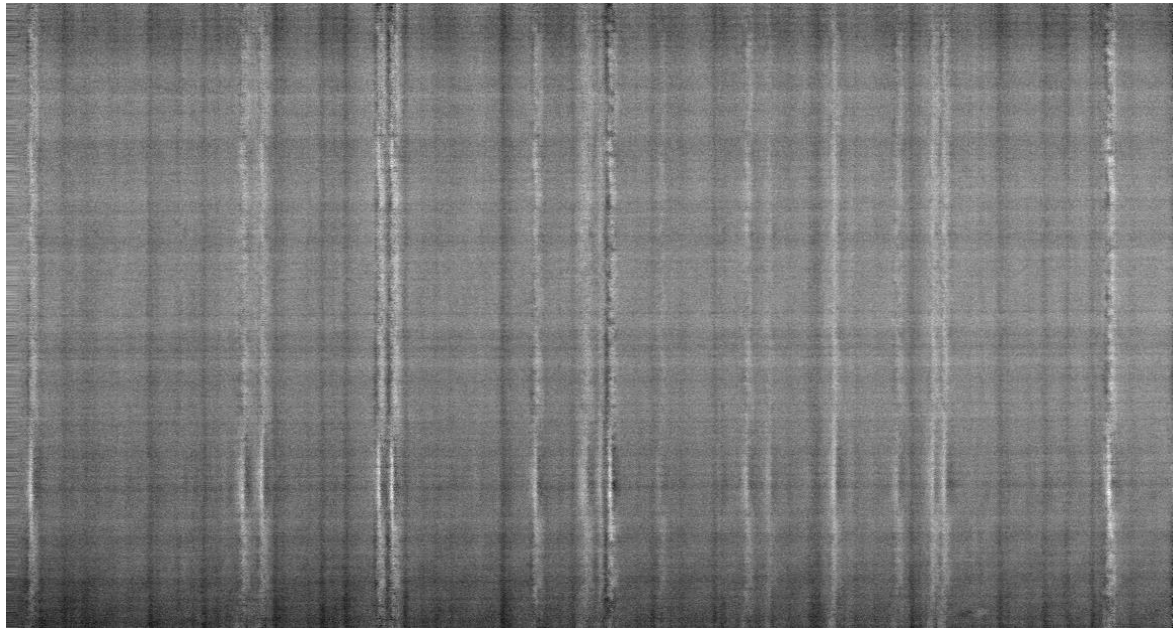

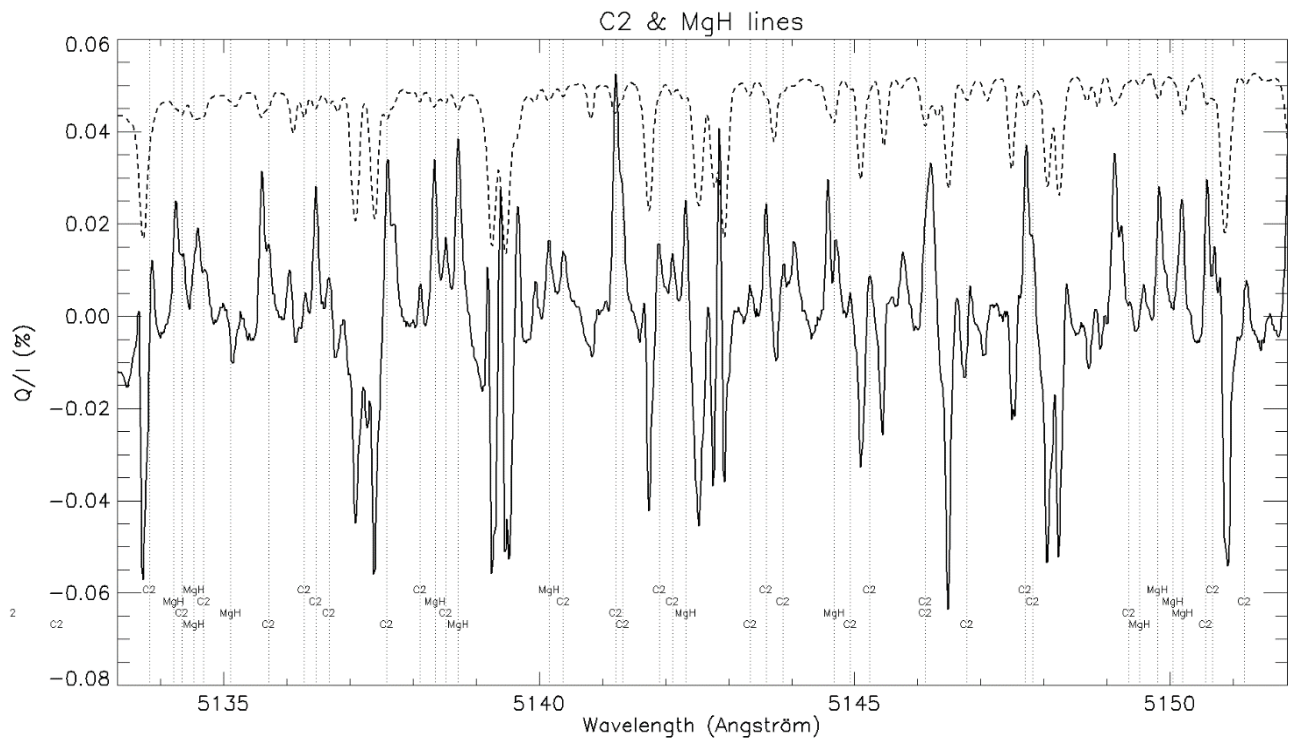

Figure 6: *$C_2$ & MgH lines in the range 5134-5152 A. Q/I along the slit at top before summing along the slit, near **north pole**. Bottom: solid line, Q/I; dashed line, I.*

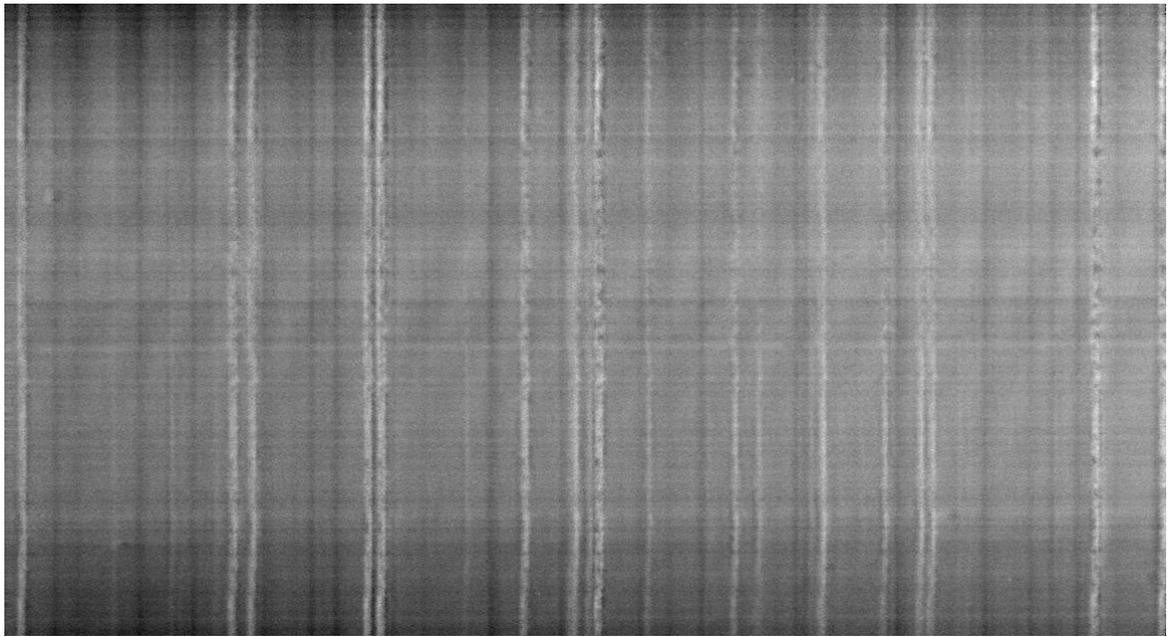
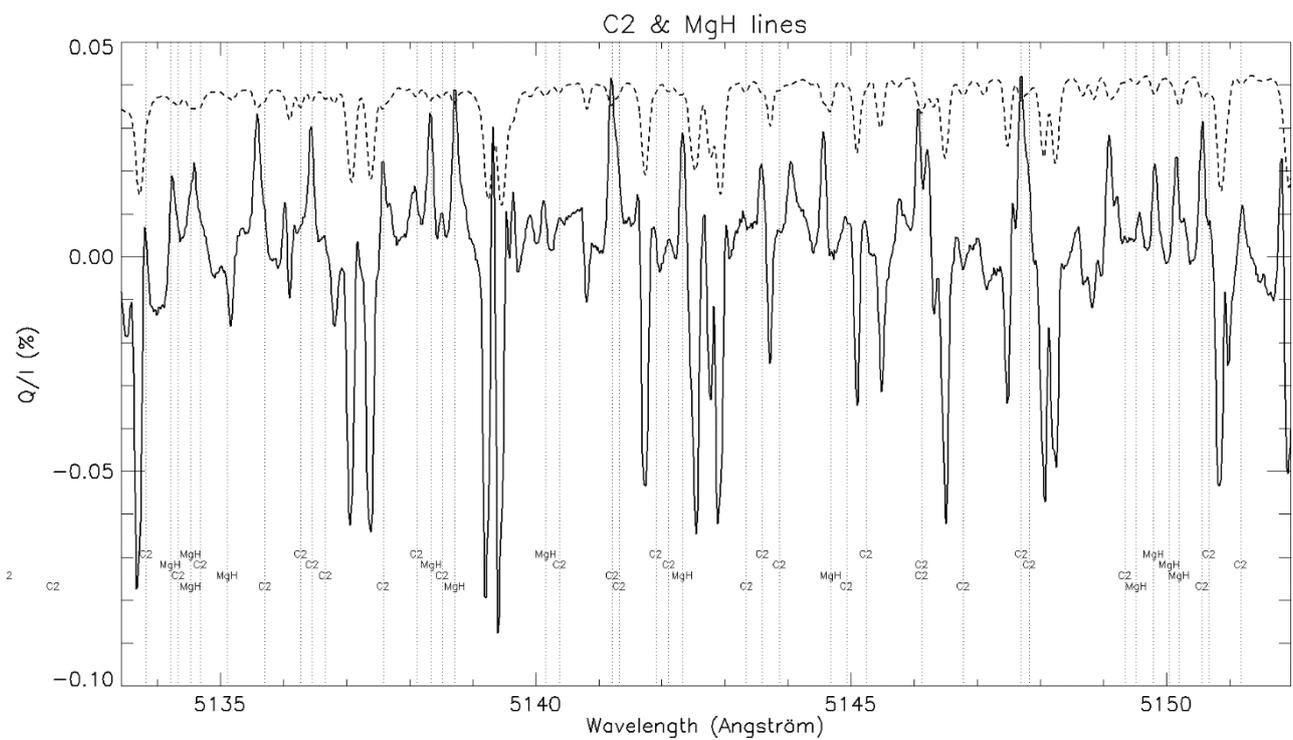

*Figure 7: $C_2$ & MgH lines in the range 5134-5152 A. Q/I along the slit at top before summing along the slit, near **south** pole. Bottom: Solid line, Q/I; dashed line, I.*

## II – 3 - The Hydrogen Hα 6563 A line (figure 8)

For Hα line, 4 sets are available (11" above the limb, 4" above the limb, at the limb and 4" below the limb, on the disk); the slit was orthogonal to the limb, and we summed over 4" only to increase the SNR instead of summing along the entire slit. Results are displayed in figure 8. The polarization of this line appears easily in the second solar spectrum.

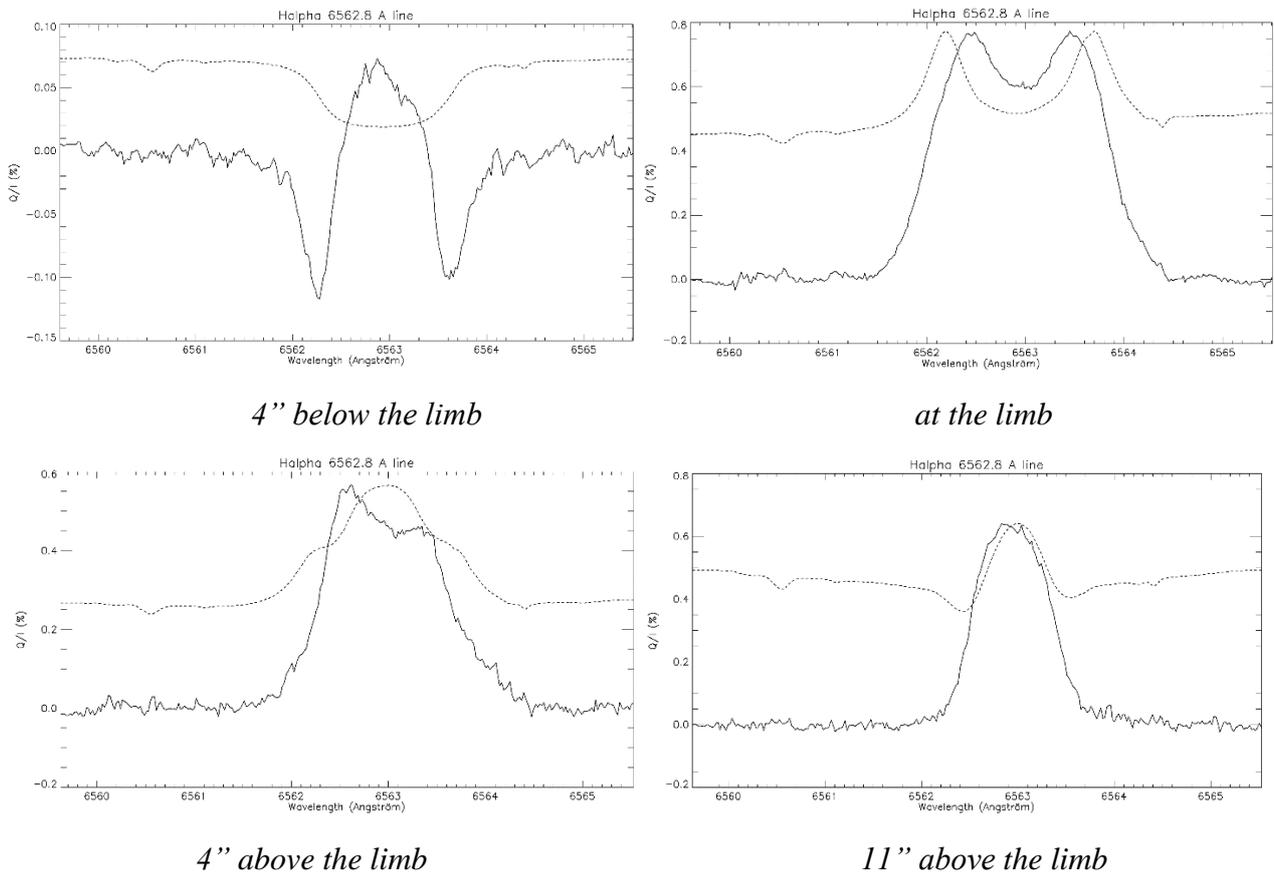

*4" below the limb*                                            *at the limb*

*4" above the limb*                                            *11" above the limb*

*Figure 8: Hα line at various locations: 4" below the limb (the line is in absorption); at the limb (the wings are in emission, but the core is not yet); at 4" above the limb (line in emission) and 11" above the limb (line in emission). Solid line: Q/I; dashed line: I.*

## II – 4 – Helium D3 & Sodium NaD2/NaD1 lines (figures 9 & 10)

HeD3 5876 was observed alone (figure 9) and together with NaD2 5890 and NaD1 5896 (figure 10). HeD3 is not visible on the disk but has a polarization signature near the limb.

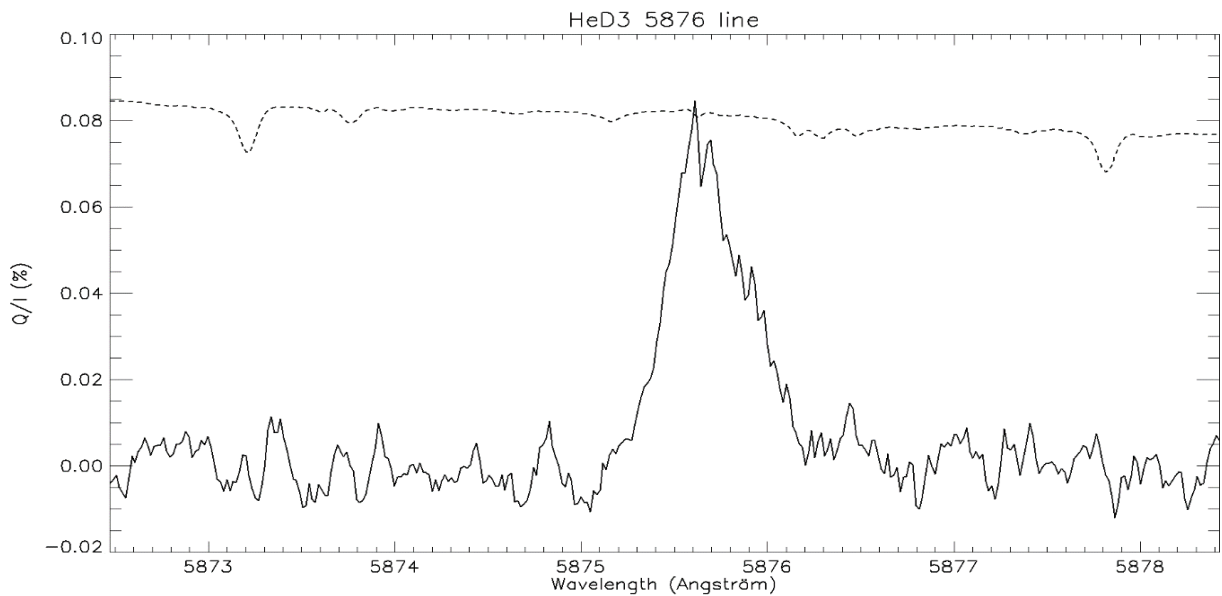

*Figure 9: He D3 line at 10" below the limb (the line has a signature in polarization, but is not visible in intensity). Solid line: Q/I; dashed line: I.*

The slit was orthogonal to the limb and data were integrated over 2" only to produce the figure 10.

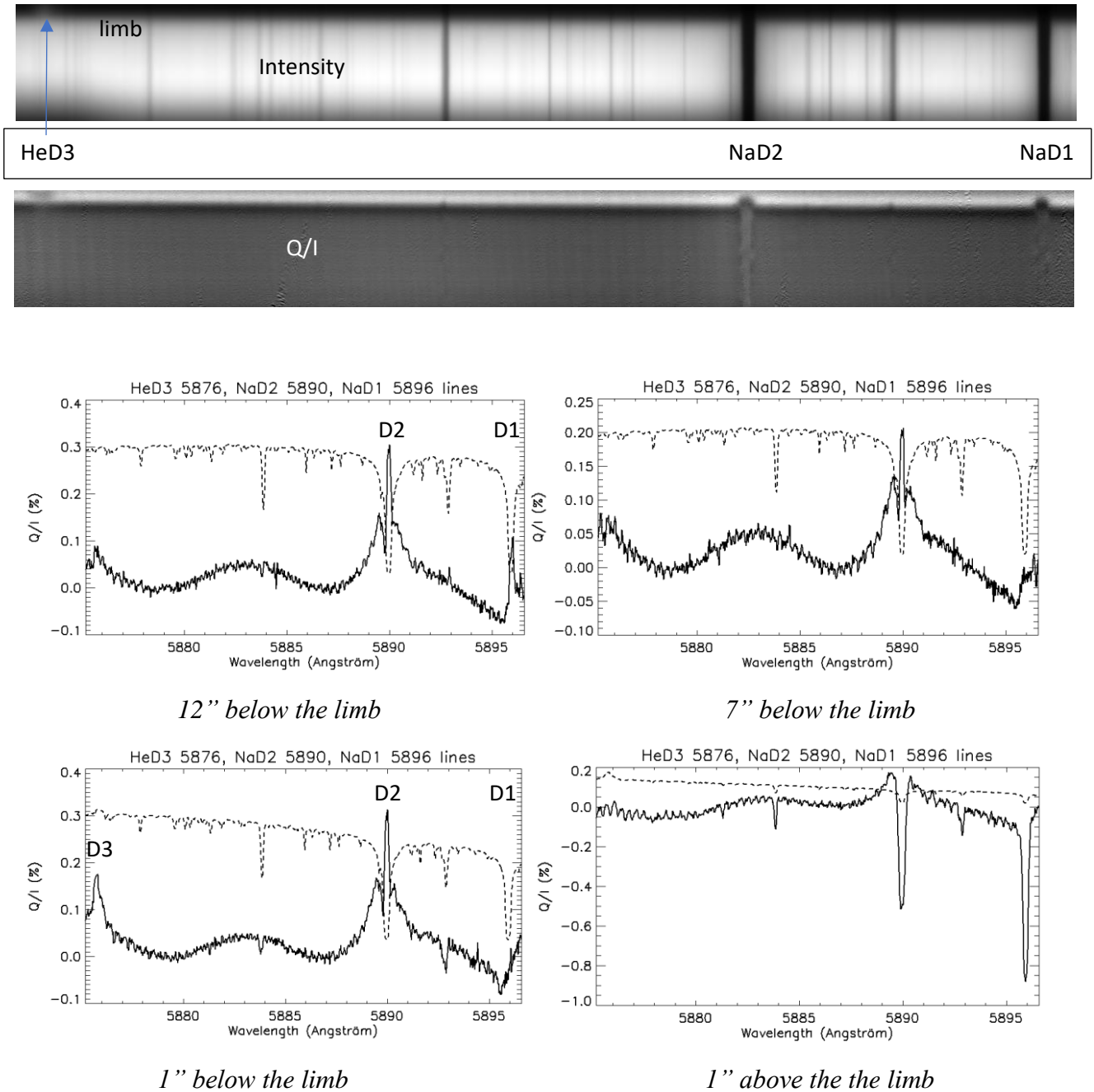

*12" below the limb*   *7" below the limb*

*1" below the limb*   *1" above the the limb*

*Figure 10: He D3, NaD2 and NaD1 lines at various distances of the limb). Below the limb, on the disk, HeD3 does not appear in intensity, but is polarized. Above the limb, HeD3 is visible in intensity in the chromosphere, but NaD1 and D2 completely vanish in the scattered light. There is a large oscillation between D3 and D2 which is parasitic and caused by an optical default over the 20 A spectral range. Solid line: Q/I; dashed line: I.*

## II – 5 - NaD1 5896 and NaD2 5890 lines (figures 11 & 12)

NaD2 and NaD1 have an enigmatic behaviour that was studied by Stenflo *et al* (2000b); NaD1 is much weaker in polarization than NaD1 (figure 11); NaD2 triplet peak is well resolved by our observation, but this is not the case of NaD1. Two series were obtained for NaD2 that are shown below (figure 12).

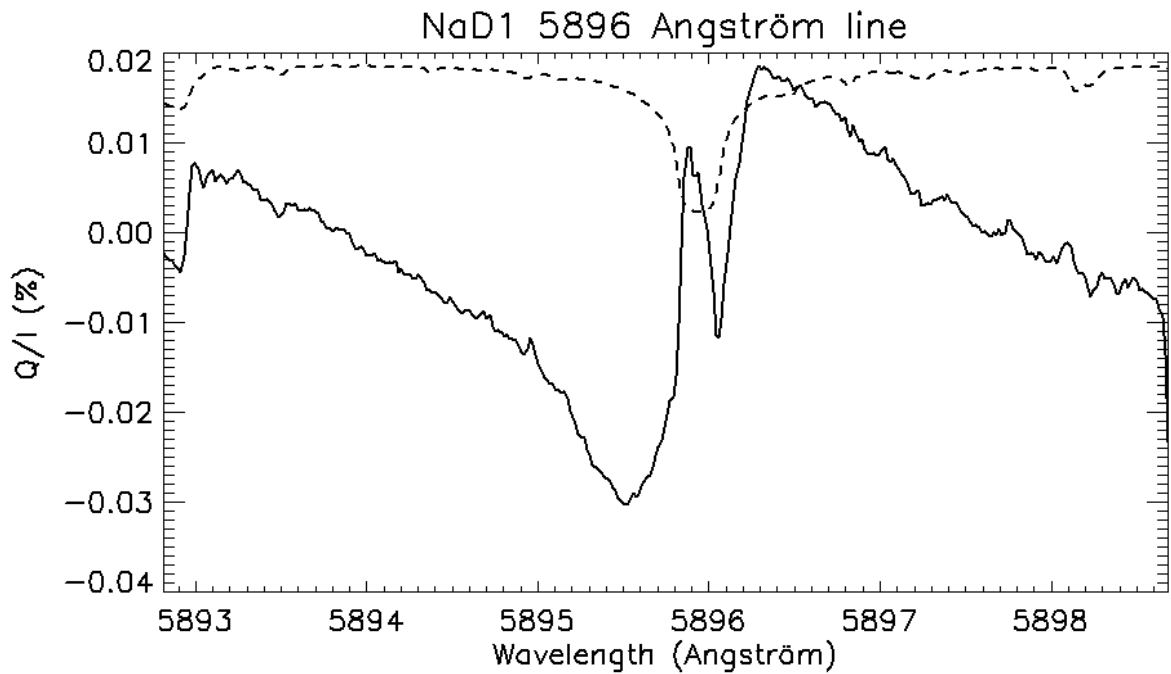

*Figure 11: Na D1 5896 A line. Solid line: Q/I; dashed line: I.*

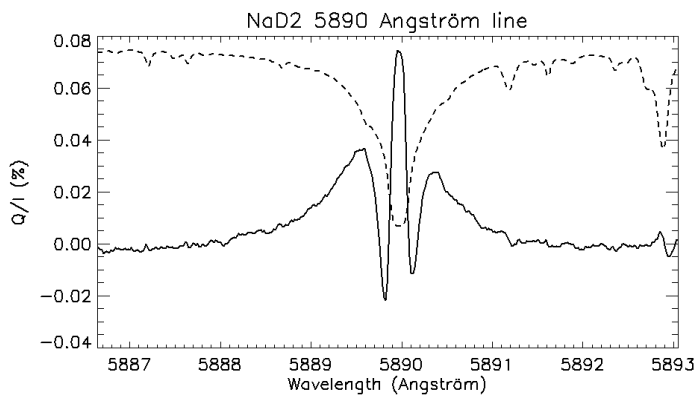
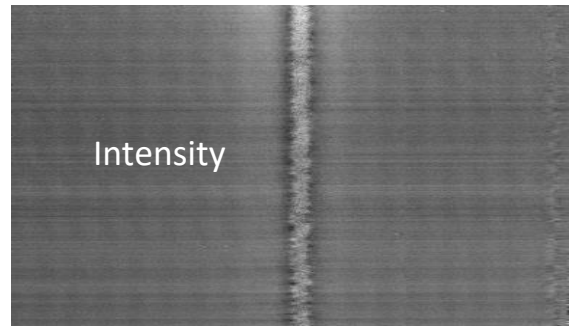
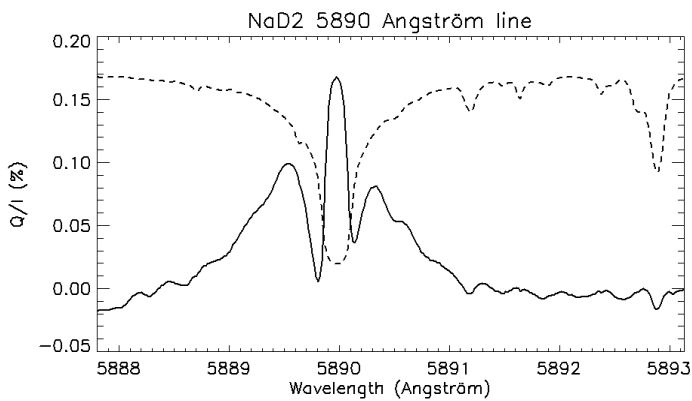
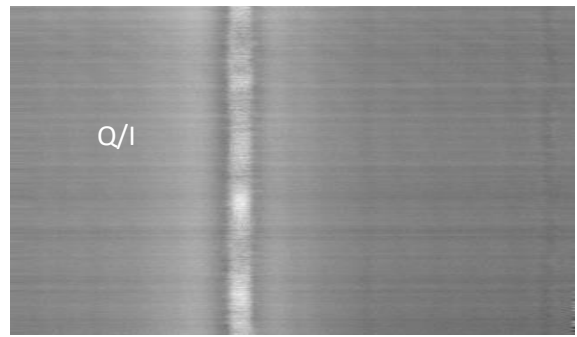

*Figure 12: Na D2 5890 A line. Left: solid line: Q/I; dashed line: I. Right: intensity and Q/I before integration along the slit.*

## II – 6 - Lithium D1/D2 6708 lines (figure 13)

Li D1/D2 lines are very close and faint in the intensity spectrum, and they were investigated by Stenflo et al (2000a). Two series of observations (figure 13) were performed at Pic du Midi, showing a small Q/I peak at Li 6708 after integration along the slit, despite of interference fringes.

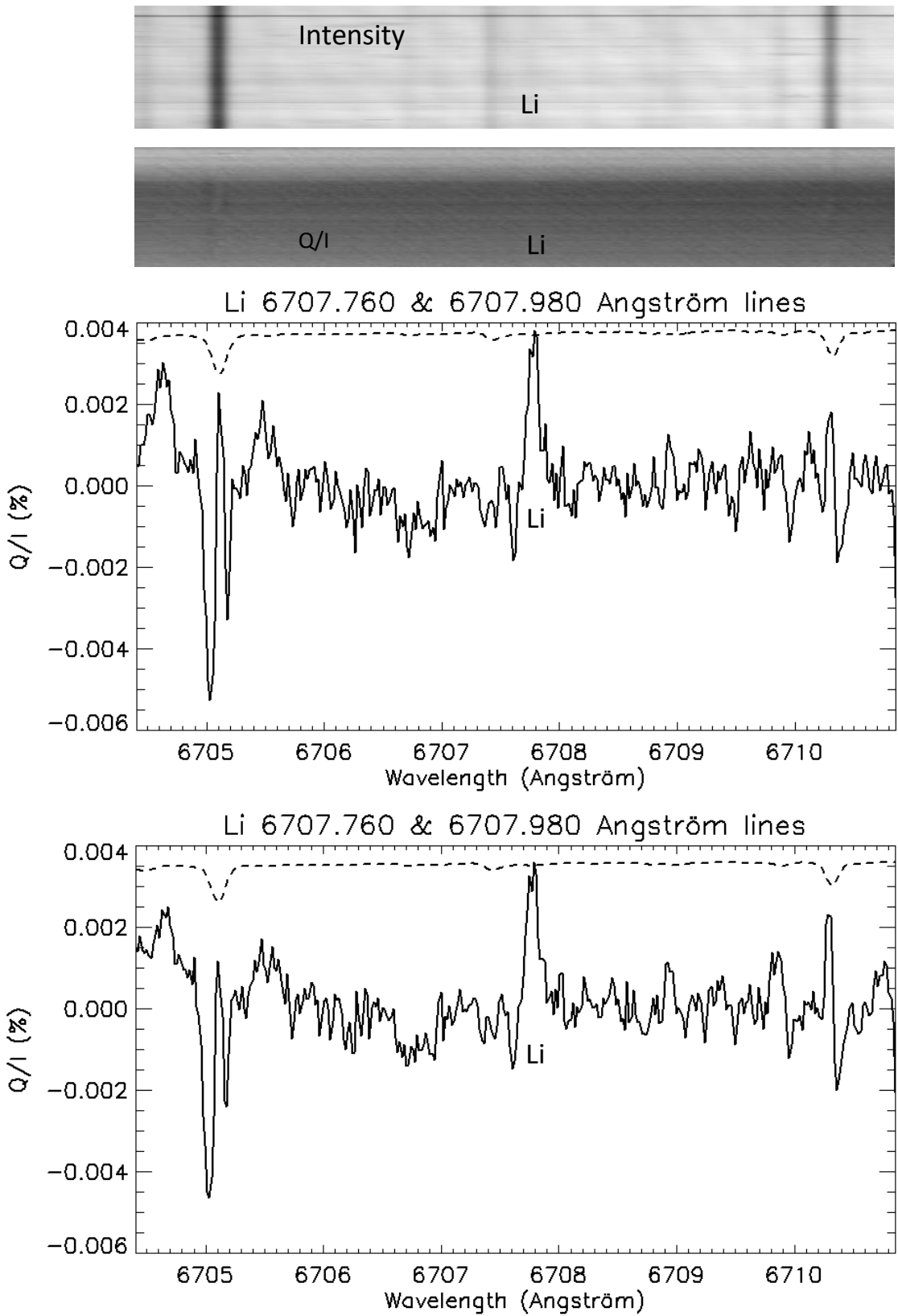

*Figure 13: Li 6708 A D1/D2 lines. Top: intensity and Q/I spectra. Bottom: solid line: Q/I; dashed line: I.*

**II – 7 - Scandium ScII 4247 A line (figure 14)**

The hyperfine splitting in Stokes Q/I of ScII was noticed by Stenflo (2000). It is well evidenced by our observations of figure 14.

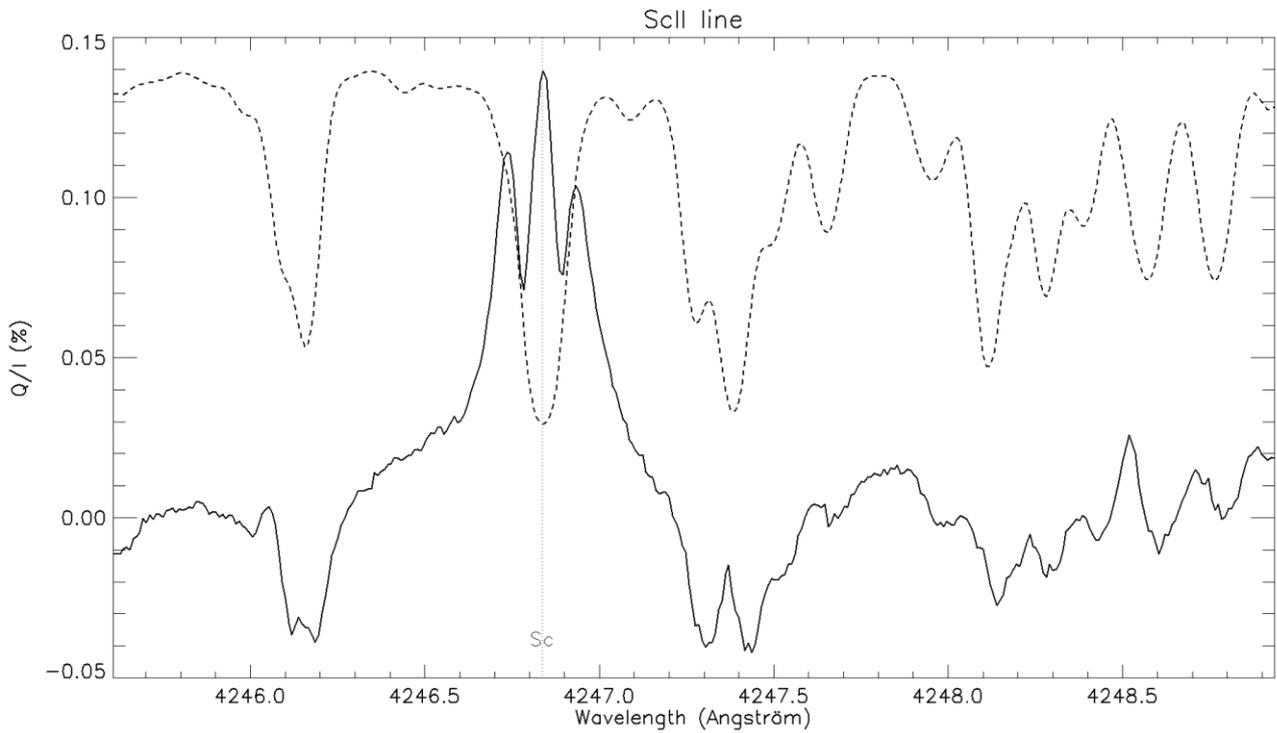

*Figure 14: ScII 4247 A line. Solid line: Q/I; dashed line: I.*

## II – 8 - Titanium lines around 4536, 4642, 4742, 4758, 4760, 4778 Angström (figures 15 to 23)

Titanium lines are very numerous in the blue/green part of the solar spectrum. We observed many spectral ranges in which we saw polarized and unpolarized lines of Titanium (figures 15 to 23). All observations were performed on the disk at 10" of the solar limb with the slit parallel to the limb; pixels were integrated along the slit.

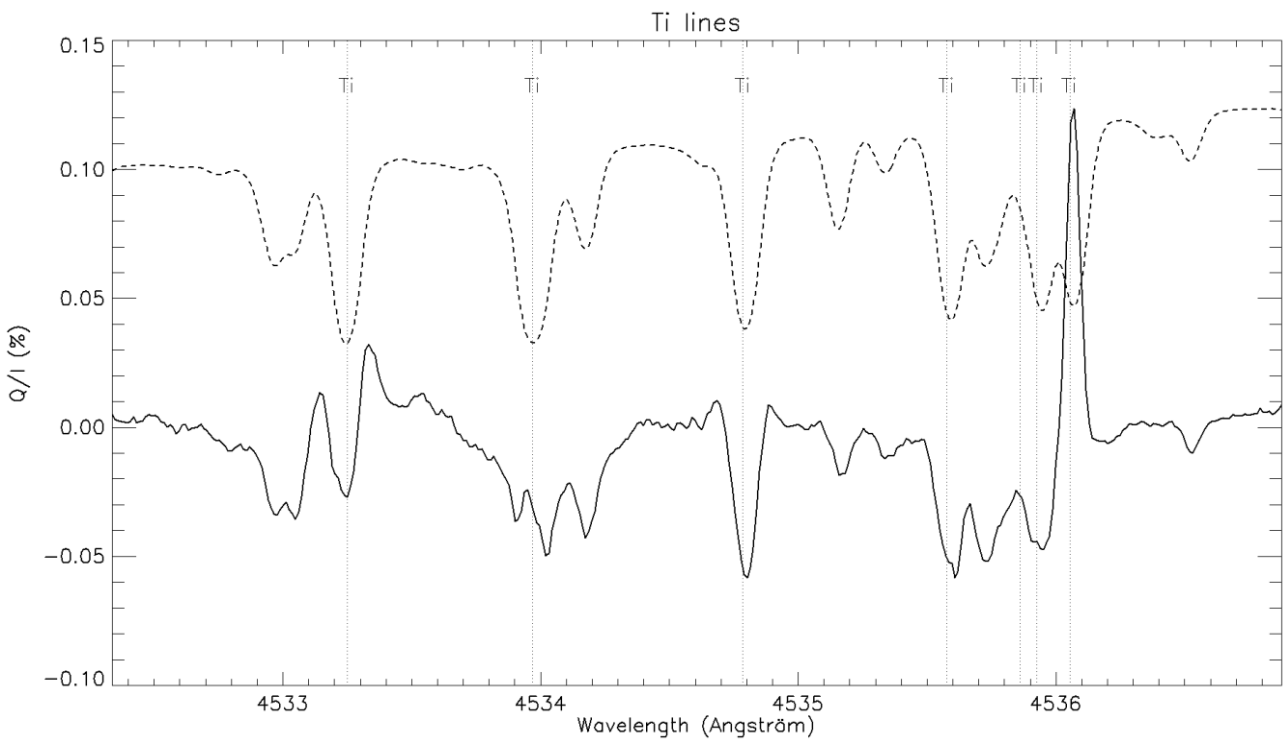

*Figure 15: TiI lines in the range 4532-4537 A. Solid line: Q/I; dashed line: I.*

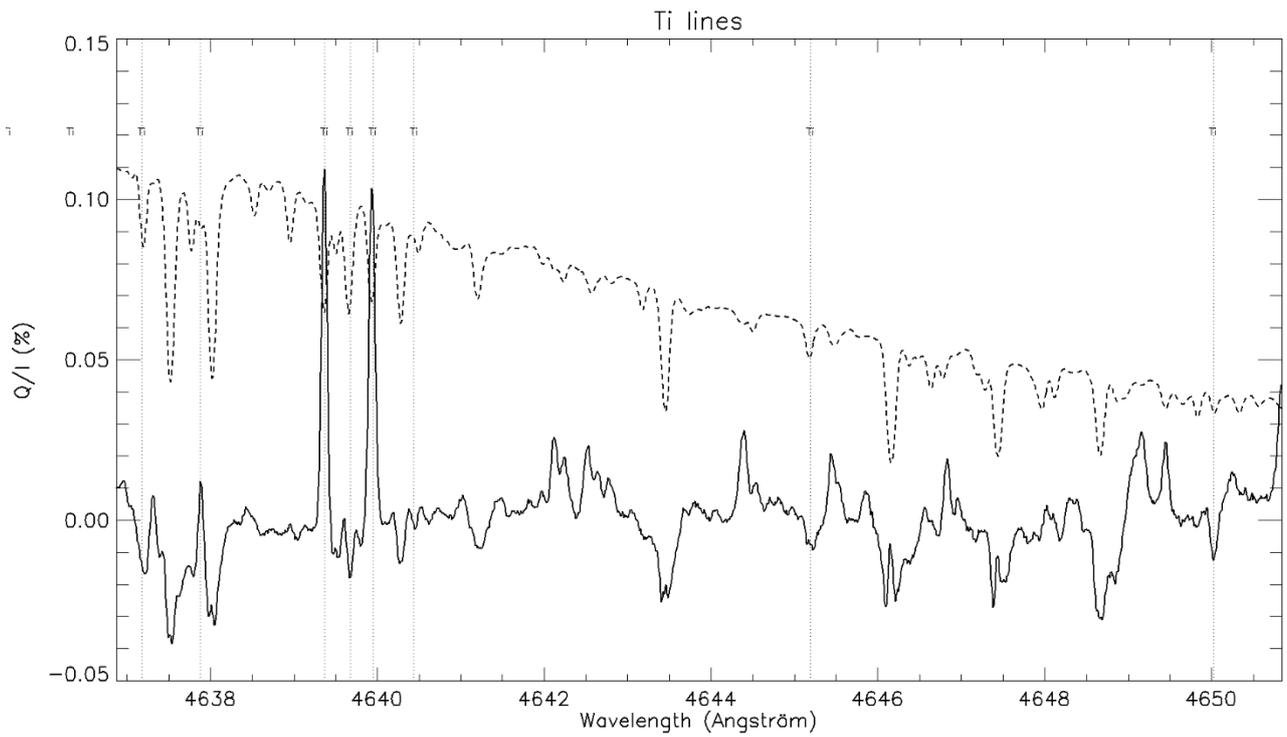

*Figure 16: TiI lines in the range 4638-4650 A. Solid line: Q/I; dashed line: I. The decreasing intensity with wavelength is explained by the fact that we are in the wing of the selecting order filter, with decreasing transmission with wavelength.*

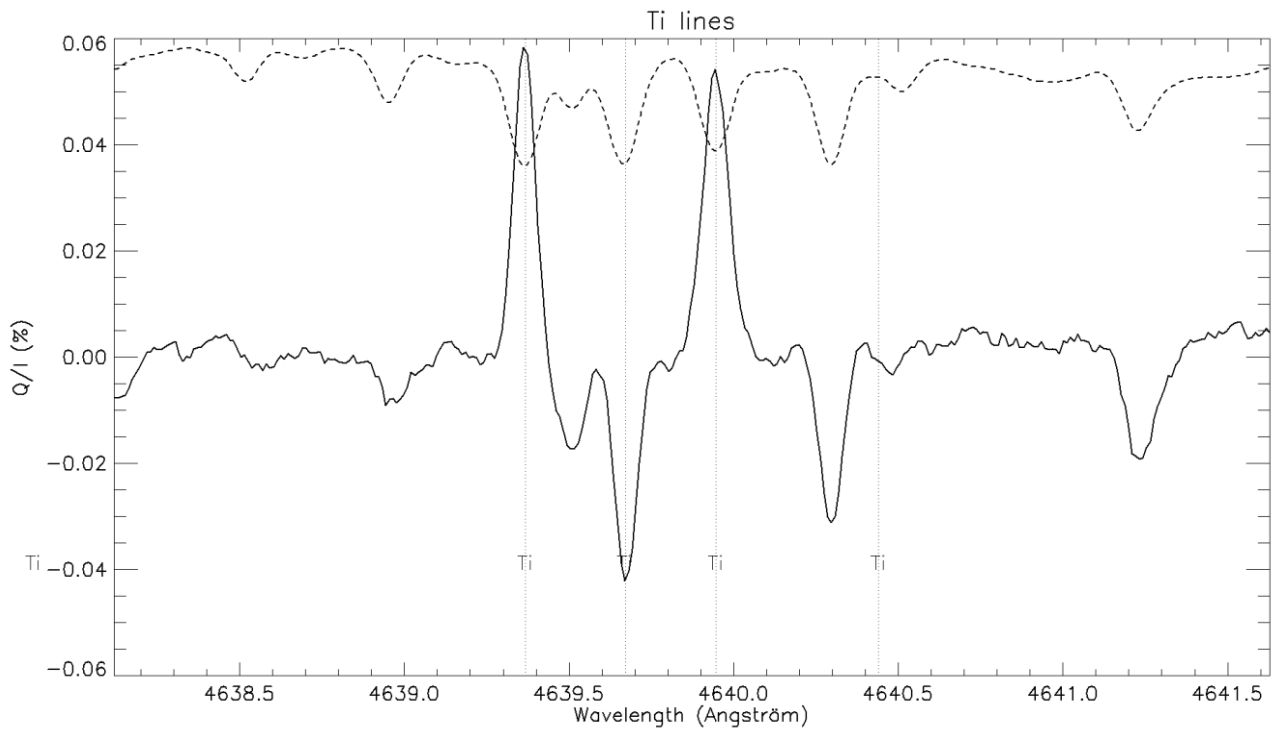

*Figure 17: TiI lines in the range 4638-4641 A. Solid line: Q/I; dashed line: I. It is a second observation of the spectral domain included in the left part of figure 16.*

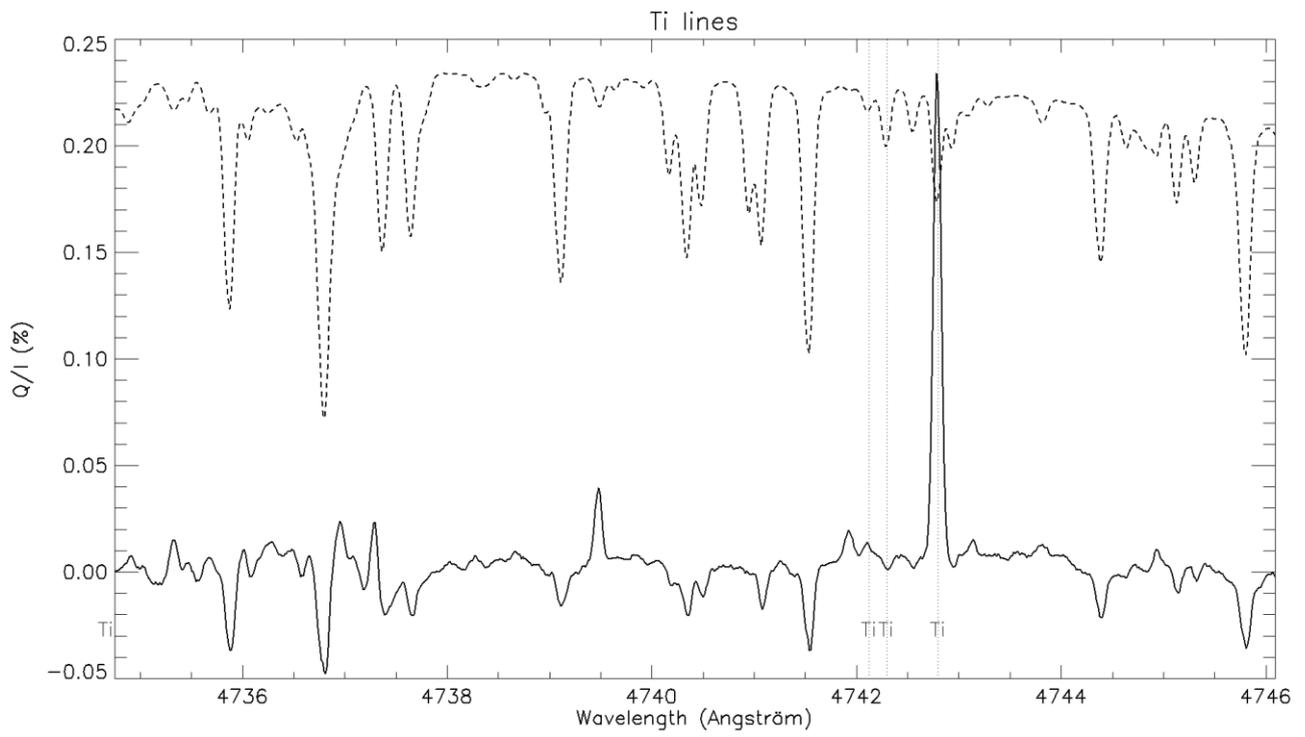

*Figure 18: TiI lines in the range 4735-4746 A. Solid line: Q/I; dashed line: I.*

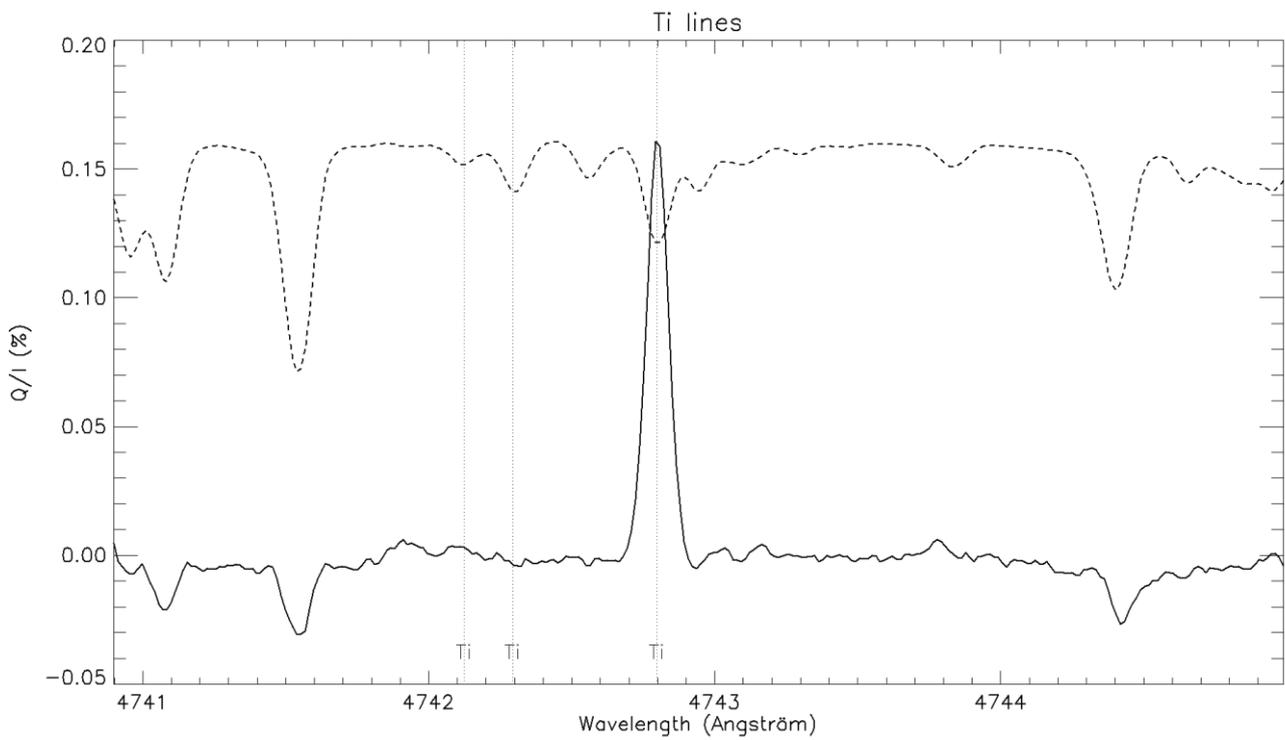

*Figure 19: TiI lines in the range 4741-4745 A. Solid line: Q/I; dashed line: I. It is a second observation of the spectral domain included in the right part of figure 18.*

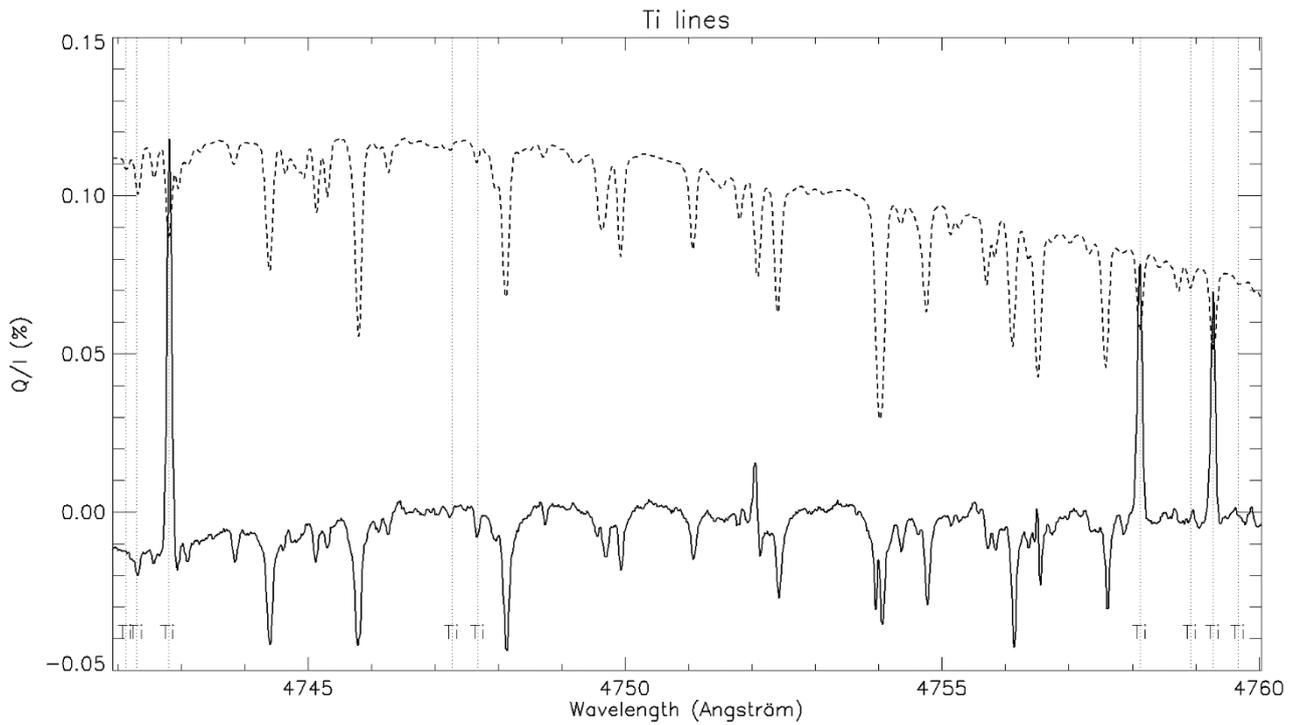

*Figure 20: TiI lines in the range 4742-4760 A. Solid line: Q/I; dashed line: I. The decreasing intensity with wavelength is explained by the fact that we are in the wing of the selecting order filter, with decreasing transmission with wavelength.*

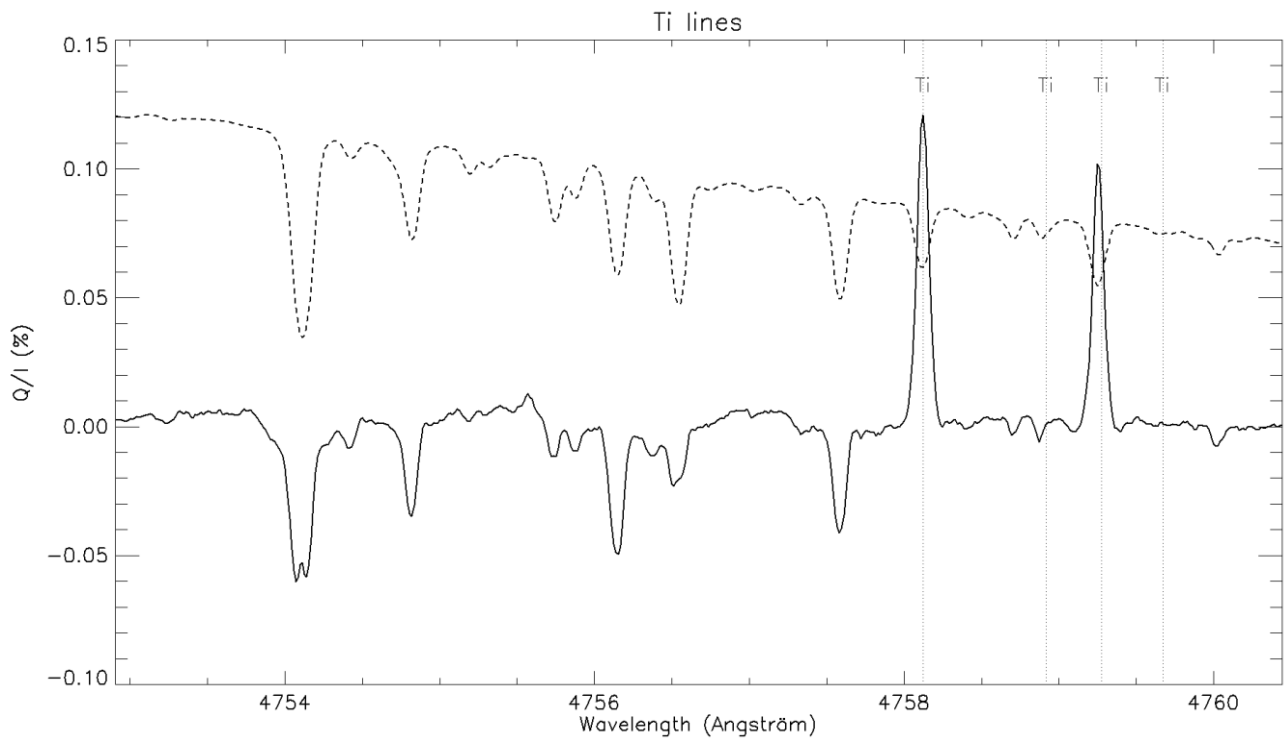

*Figure 21: TiI lines in the range 4753-4760 A. Solid line: Q/I; dashed line: I. The decreasing intensity with wavelength is explained by the fact that we are in the wing of the selecting order filter, with decreasing transmission with wavelength. It is a second observation of the spectral domain included in the right part of figure 20.*

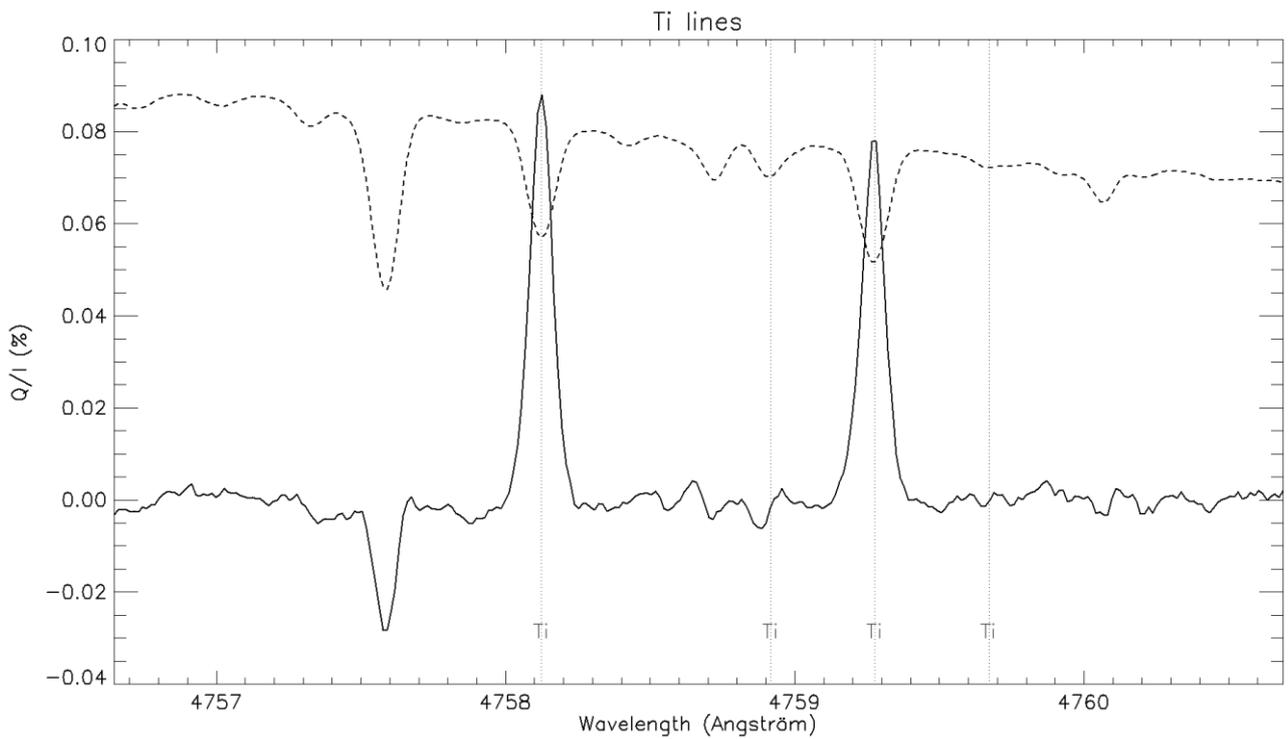

*Figure 22: TiI lines in the range 4757-4760 A. Solid line: Q/I; dashed line: I. The decreasing intensity with wavelength is explained by the fact that we are in the wing of the selecting order filter, with decreasing transmission. It is a second observation of the spectral domain included in the right part of figure 21.*

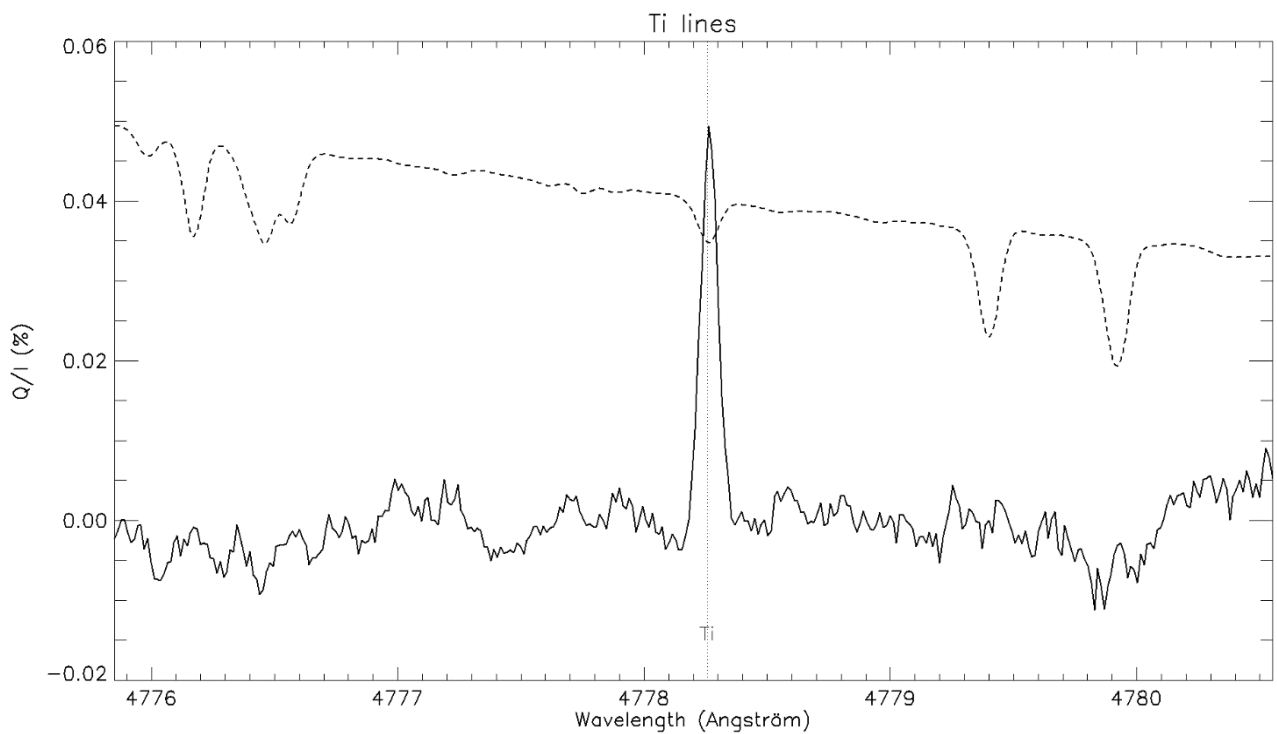

*Figure 23: TiI lines in the range 4776-4781 A. Solid line: Q/I; dashed line: I. The decreasing intensity with wavelength is explained by the fact that we are in the wing of the selecting order filter, with decreasing transmission.*

**III – The dataset**

The present data are available on line in FITS format at https://doi.org/10.57745/FEXIZ0

In particular, files "int.fits" (intensity), "q_over-i.fits" and "wl.fits" correspond exacty to the figures shown above for each line of the dataset.

**IV - Conclusion**

The present data were obtained with a liquid crystal polarimeter and a classical interline CCD camera running at 5-10 Hz cadence. Stokes combination I+Q and I-Q were obtained sequentially, 16 couples were integrated in real time within a few seconds to achieve the SNR of 500, in order to form two elementary 16 bit frames ; then hundreds of such frames were got in sequence and added, in order to increase the SNR to 10000; and finally, for some lines, pixels along the spectrograph slit were summed to reach a maximum SNR of $10^5$, allowing the measurement of faint linear polarization rates Q/I. Our data are not perfect and suffer from optical defaults such as fringe networks and field non uniformity, that are not fully corrected and limit the detection of very weak polarizations. However, our sensitivity is better than the one of the atlas of the second solar spectrum and corroborates deep polarimetric observations made by the Zürich team on specific lines with ZIMPOL.